\documentclass[preprint,12pt]{elsarticle}

\usepackage{graphicx}
\usepackage{dcolumn}
\usepackage{bm}
\usepackage{amsmath}
\usepackage{amssymb}
\usepackage{amsfonts}
\usepackage{epsfig}
\begin{document}
\begin{frontmatter}

\title{Transition from the self-organized to the driven dynamical clusters}
\author{Aradhana Singh and Sarika Jalan}
\address{Complex Systems Lab, Indian Institute of Technology Indore,
IET-DAVV Campus Khandwa Road, Indore 452017, India}
\date{\today}

\begin{abstract}
We study mechanism of formation of synchronized clusters in coupled maps on 
networks with various connection architecture. The nodes in a cluster are self synchronized or driven 
synchronized, based on the coupling strength and underlying network structures. Smaller coupling strength
region shows driven clusters independent of the network rewiring strategies, where as
larger coupling strength region shows the 
transition from the self-organized cluster to the driven cluster as network connections 
are rewired to the bi-partite type. Lyapunov function analysis is performed to 
understand
the dynamical origin of cluster formation. The results provide insights into the relationship 
between the topological clusters which are based on the direct connections between the 
nodes, and the dynamical clusters which are based on the functional behavior of these 
nodes.

\end{abstract} 
\end{frontmatter} 

\section{Introduction} 

Networks of dynamical units have been used to model everything from earthquakes to 
ecosystems, neurons to financial market 
\cite{Winfree,Kuramoto-book,Wiesenfeld,May,Strogatz,Barabasi,Boccaletti}. Studying the 
impact of complex network topology \cite{Strogatz,Barabasi} on the dynamical 
processes is of fundamental importance for understanding the functioning of these real 
world systems. Synchronization is one of the most fascinating phenomenon 
exhibited by such nonlinear dynamical units interacting with each other 
\cite{book-syn,PhyRep-syn}. This phenomenon has been intensively investigated as a 
universal concept in many disciplines, particularly in neuro-sciences \cite{syn-neuro} 
and communication \cite{syn-com}. Recently synchronization has been shown to play a 
crucial role in the species extinction \cite{REA-species}.  Keeping such applications 
in mind, a great deal of work is devoted to build up the strategies for the network 
connection architecture which could synchronize the dynamical units on them 
\cite{MSF}. Most of these works have considered global synchronized states spanning all 
nodes in a network.

In the past years focus of coupled dynamics research got shifted to the (phase) synchronized 
clusters \cite{PhaseSyn} or dynamical clusters instead of a global synchronized state. Dynamical 
clusters are defined based on some kind of coherent time evolution of some units which is different 
than the evolution of other units. Dynamical clusters have been the part of coupled dynamics 
research since the seminal work of Kuramoto which mainly focused on the transition to the global 
synchronized state \cite{Kuramoto}. Before the transition, \cite{Kuramoto} reports the 
phase-synchronized clusters, where clusters are distinguished based on the frequencies of dynamical 
units. Kaneko's coupled maps model also talks about the glassy phase where nodes are partially 
synchronized \cite{Kaneko}. Kuramoto's coupled oscillator and Kaneko's coupled maps models assumed 
either nearest neighbor connections or all to all connections between the nodes. Later on with the 
spurt of complex network research, these types of coupled dynamics models had been investigated 
with units having local and nonlocal connections, and also show partial synchronized state with 
dynamical clusters \cite{DynClus-net}. These works mainly concentrated on the finding of partial 
synchronized state, and did not pay much attention to the coupling configurations. However, 
networks modeling real world, have some structures, hence it was very natural to imagine a relation 
between the connection architecture and the dynamical clusters. It could be rephrased as follows: 
do connections in the dynamical clusters have some specific features ? Manrubia and Mikhalov in 
\cite{Manrubia} found that the connections in dynamical clusters do have some particular 
configuration, i.e. elements from a cluster generally having more connections inside the cluster 
than with the elements from other dynamical clusters. Though this was a very significant 
observation and more importantly presented the concept of relation between dynamical cluster and 
the underlying network, their observation was based on the dynamical units coupled with some 
particular coupling strength and having random coupling architecture. The detailed investigations 
of coupled dynamics on various networks, done by Jalan and Amritkar \cite{SJ-prl1}, revealed that 
there exists two different mechanisms of the formation of the dynamical clusters; (a) 
self-organized clusters where couplings are intra-cluster type, (b) driven-synchronized where 
coupling are inter-cluster type. The main result of the Ref.\cite{SJ-prl1} is that the coupled 
dynamical units may get synchronized because of inter-cluster or intra- cluster couplings. Whereas 
self-organized mechanism has more commonly been thought as a reason of synchronization from seminal 
work by Kaneko \cite{Kaneko} to recent work by several scientists \cite{Manrubia,Guilera_pre2006, 
Changsong_pre2010}, \cite{SJ-prl1} identified a new mechanism of cluster formation that is driven 
synchronization and did extensive studies of the two mechanisms of cluster formation for various 
networks \cite{SJ-prl1,SJ-pre1,SJ-analytic}. Recently self and driven-synchronization has been 
investigated in the relevance of brain cortical networks \cite{Changsong2011}.

Real world networks have community or module structure \cite{Newman,modular}, where modules are the 
division of network nodes into various groups within which the network connections are dense, but 
between which they are sparser. The modularity concepts assumes that system functionality can be 
partitioned into a collection of modules and each module is a discrete entity of several components 
performing an identifiable task, separable from the functions of other modules \cite{modular}. In 
the light of this finding the concept of relation between dynamical clusters and the network 
architecture \cite{Manrubia,SJ-prl1} can be rephrased; do dynamical clusters reveal organization or 
module structure of the underlying network, and what functional clusters would be preferred for a 
given network architecture ? Investigation in this direction \cite{ClusModule-Zhou, 
ClusModule-ABGuilera} showed that the dynamical units reveal the hierarchical organization of the 
synchronization behavior with respect to the collective dynamics of the network. Furthermore Zhou 
et. al. used synchronized clusters to unveil the functional connectivity in the complex brain 
networks \cite{Brain-Zhou}. These works exploited the synchronous behavior of dynamical units to 
detect the hierarchical organization, on the other hand M. Timme concentrated on the disordered 
state to understand the relation between dynamical behavior and the network topology 
\cite{ClusModule-Timme}. Recently phenomenon of driven synchronization is observed for three 
stochastic oscillators which are coupled with delay \cite{amitabha}.

In this paper we highlight that there is a great difference in the process leading to the 
synchronization and synchronized cluster state, depending upon the particular coupling 
configuration. We show that these two ways of the construction of the dynamical clusters depend 
highly on the underlying network architecture. Self-organized clusters are formed for the networks 
whose network architecture is such that it can clearly be divided into different sub-communities, 
whereas for the bi-partite networks driven clusters are formed. Mixed clusters, i.e. some are 
self-organized and some are driven organized, are formed for the network architecture which lie in 
the middle of these two extreme ideal network structure. We use Lyapunov function to understand the 
origin of these two types of clusters.

The paper is organized as follows. After the introductory section describing the current 
understanding and importance of dynamical cluster research, we introduce general model of coupled 
dynamics in section II. Section III consists the definition of dynamical clusters and measures 
quantifying the behavior of the clusters. We explain network rewiring strategy in section IV and 
also present results for the dynamical behavior at various step of the network rewiring. Section V 
considers Lyapunov function analysis. Section VI concludes the paper with the possible future 
directions.

\section{Coupled dynamics on Networks}
Consider a network of $N$ nodes and $N_c$ connections. Let each node of the network
be assigned a dynamical variable $x_i, i = 1,2, \hdots N$. Evolution of the
dynamical variable is written as \cite{Kaneko}
\begin{equation}
x_i(t+1) = f(x_i(t)) + \frac{\varepsilon}{k_i} \sum C_{ij} [g(x_j(t)) - g(x_i(t))]
\label{cml}
\end{equation}
where $x_i(t)$ is the dynamical variable of the $i$th node at the $t$th time step, and 
$C$ is the adjacency matrix with elements taking values $one$ and $zero$ depending 
upon whether $i$ and $j$ are connected or not, $\varepsilon \in [0,1]$. 
Matrix $C$ is symmetric matrix representing
undirected network, and $k_i = \sum_j C_{ij}$ is the degree 
of node $i$. Function $f(x)$ defines the local nonlinear map and function 
$g(x)$ defines the nature of coupling between nodes. In this paper we present  
results for the local dynamics given by logistic map $f(x) = \mu x (1-x)$ in the 
chaotic regime and $g(x) = f(x)$. This coupled maps model is studied extensively to 
get insight 
into the behavior of real systems \cite{Kaneko,rev-CM}. We take the value of $\mu=4$, for 
which individual un-coupled unit shows chaotic behavior with the value of Lyapunov exponent 
being $\ln 2$. As an effect of coupling the coupled dynamics \ref{cml} shows various different 
kinds of coherent behavior, such as synchronization \cite{Kaneko}, 
phase-synchronization \cite{PhaseSyn,SJ-prl1,SJ-pre1,SJ-physicaA} and other large scale macroscopic 
coherence \cite{SJ-sym} depending upon the coupling architecture and the coupling 
strength. 
In this paper we consider synchronization $x_i(t) = x_j(t)$ for all $i,j$ and all initial conditions,
and phase-synchronization as considered in \cite{phase-prl1998,SJ-prl1}. Let $n_i$ and $n_j$ denote the 
number of times the variables $x_i(t)$ and $x_j(t)$ , $t = 1, 2, . . . , \tau$ for the nodes $i$ and $j$ 
show local minima during the time interval $\tau$ . Let $n_{ij}$ denote the number of times these local 
minima match with each other. Phase distance between nodes $i$ and $j$ is then given by
$d_{ij} = 1 - 2n_{ij}/(n_i + n_j)$. Node $i$ and $j$ are phase synchronized when $d_{ij}=0$. 
In a dynamical cluster, all the pairs of nodes are (phase) synchronized.

\section{Synchronized clusters} 
Various types of dynamical clusters based on the 
couplings between the nodes are defined as:
(a) Self-organized cluster: If nodes of a cluster are synchronized because of 
intra-cluster couplings, the cluster is referred as self-organized cluster, and (b) driven cluster: 
If nodes of a cluster are synchronized because of inter-cluster
connections, the cluster is referred as driven cluster.
Quantitative measure of above two types of clusters are given by the following
two quantities \cite{SJ-prl1,SJ-pre1};
\begin{eqnarray}
f_{intra} = \frac{N_{intra}}{N_c} \nonumber \\
f_{inter} = \frac{N_{inter}}{N_c} \nonumber
\end{eqnarray}
where $N_{intra}$ and $N_{inter}$ are the number of intra- and inter cluster 
connections, respectively. $f_{inter}$ and $f_{intra}$ would take values between $zero$ and $one$.
For the ideal driven or ideal self-organized case one of the measure would take value
$one$. According to this definition $f_{inter}$ and $f_{intra}$ both would be zero
when none of the nodes form cluster. The intermediate values of these two quantities represent
the situation where some of the nodes are of driven type and some of them are of
self-organized type, or it may correspond to the case when some clusters have mixed
connections, i.e. some of connections are intra-cluster type whereas some are
inter-cluster type.  $f_{inter} +
f_{inter} = 1$ represents the situation when all the nodes form one or many
synchronized clusters and $0< f_{intra} + f_{inter} < 1$ corresponds to the case
when some nodes do not belong to any cluster and evolve independently.
In the dominant self-organized clusters number of intra-cluster connections
are much greater than the inter-cluster connections, while in the dominant driven clusters reverse
is true. Mixed clusters corresponds to the case when 
number of inter and intra-cluster connections are roughly
equal. Note that while counting $f_{inter}$ we do not
consider isolated nodes, only the nodes which are the part of any synchronized cluster
are considered. 

Newman's definition of modularity of a network, $Q$, is also based on the ratio of intra and inter 
connections ($Q = 1/2N_c [\sum_i e_{ii} - \sum_{ijk} e_{ij} e_{ki}] $), where $e_{ij}$ is the 
fraction of edges in the network that connect nodes in group $i$ to those in groups $j$. Modularity 
of a network is $one$ if it can be divided into subgraphs such that nodes in each subgraphs are 
connected within it, with a single connection to any of the other subgraph. At least one connection 
is required in order to keep the network connected. We would call these clusters, structural 
clusters (or community) to distinguish it from the dynamical clusters. In self-organized state, 
nodes in same community get synchronized dynamically also, while for the driven state, nodes which 
are synchronized have connections only to the other dynamical clusters.

In the first case, it is more obvious, that the nodes which are connected, tend to synchronize with 
one another. This concept was exploited in \cite{Brain-Zhou} to find out functional hierarchy in 
cat's cortex network. However, as shown in the Refs. \cite{SJ-prl1,SJ-pre1,SJ-analytic}, dynamical 
clusters could be formed without having even a single connection within the cluster ({\it ideal 
driven state}). These two types of clusters, self-organized and driven, could be observed 
irrespective of the dynamical state of the Eq.\ref{cml}, i.e. whole dynamics may lie on chaotic, 
periodic, quasiperiodic attractor depending upon the coupling strength and the connection 
architecture. Different possible states of this coupled dynamics are discussed in details in 
\cite{SJ-pre1}. This paper focuses on the relation between different types of clusters with the 
structure of the coupling matrix $C$. We particularly track the nature of dynamical cluster with 
the rewiring of network and show that depending upon the modularity of network, the type of 
dynamical clusters changes. If network is completely modular then in general self-organized 
clusters are preferred, whereas for bi-partite kind of network driven clusters are preferred.

In the following we present results for the coupled maps on networks, as described by the 
Eq.~(\ref{cml}), evolved with random initial conditions. First the initial network and the rewiring 
procedure are explained and then various results demonstrating the relation between the dynamical 
clusters and network structure are demonstrated.
\begin{figure}
\centerline{\includegraphics[width=6cm,height=5cm]{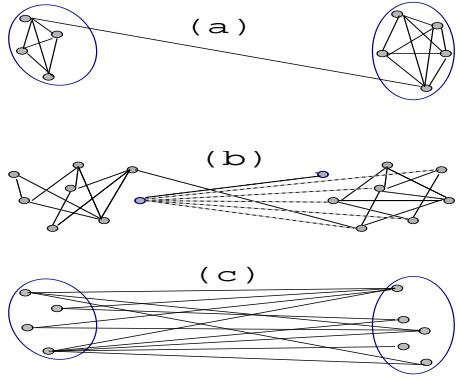}}
\caption{Pictorial representation of networks at different rewiring step. (a) Initial network with two
complete subgraphs $G_1$ and $G_2$ (b) rewired network at $n=1$, solid lines are the 
original connections and dotted lines are the new connections which a node from the 
group $G_1$ makes with the nodes of $G_2$, (c) final bi-partite graph.}
\label{FigNet}
\end{figure}

\begin{figure}
\centerline{\includegraphics[width=7.5cm,height=5cm]{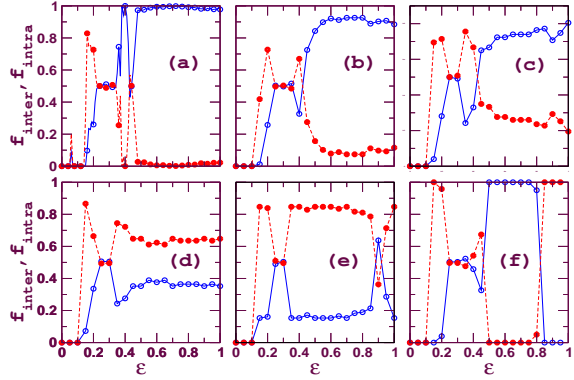}}
\caption{(Color online) Fraction of inter ($\bullet$) and intra ($\circ$) 
connections as a function of the coupling
strength $\varepsilon$. After an initial transient (about 2000 iterates) phase synchronized
clusters are studied for $\tau = 100$. The logistic map parameter is $μ = 4$, and $N_1=N_2=50$. 
(a) for the original network at rewiring step $n=0$, (b) $n =5$,
(c) $n = 10$, (d) $n=20$, (e)$n=30$, and (f)$n=50$. All figures are plotted
for the average over 20 sets of random initial conditions for the coupled dynamics.}
\label{InterIntraTime}
\end{figure}

\section{Evolution of synchronized clusters with Network rewiring} 

We start with a network of size $N$ having two complete subgraphs, $G_1$ and $G_2$ with $N_1$ and 
$N_2$ nodes respectively. Complete subgraph means, all the nodes in $G_1$ ($G_2$) are 
connected with the all other nodes in $G_1$ ($G_2$) (Fig.~\ref{FigNet}(a)). 
There exists one connection between $G_1$ and $G_2$, which is necessary to keep the 
whole graph connected. Now at each rewiring step ($n$) disconnect one node from subgraph 
$G_1$ ($G_2$) and connect it to all the nodes in the subgraph 
$G_2$ ($G_1$) (Fig.~\ref{FigNet}(b)). With this rewiring scheme, average degree of the network 
remains of the same order ($N$). After rewiring step $N_1$, all nodes in $G_1$ are connected 
with all the nodes of $G_2$, and for the case $N_1 = N_2$, the graph is complete bi-partite 
(Fig.~\ref{FigNet}(c)). At each step of the rewiring we evolve Eq.~(\ref{cml}) with  
random initial conditions and study the nature of the dynamical clusters after some initial transient. 
For small coupling strengths region ($\varepsilon < 0.5$), number of nodes forming
synchronized clusters is very small and hence in this region we consider phase-synchronized 
clusters.
For the larger coupling strengths, in general, nodes form exact synchronized clusters.
In the following we consider dynamical clusters based on the phase-synchronization.
Fig.~(\ref{InterIntraTime}) shows behavior of $f_{intra}$ and $f_{inter}$ as a function 
of coupling strength $\varepsilon$ for the various steps of the network rewiring 
according to the above strategy.  

\begin{figure}
\centerline{\includegraphics[width=7.5cm,height=5cm]{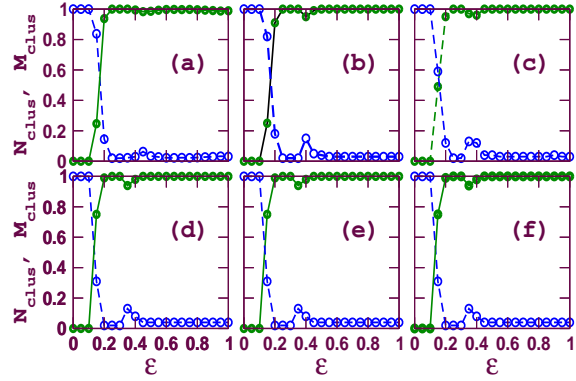}}
\caption{(Color online) The fraction of number of clusters $M_{clus} (\circ)$ and the fraction of the 
number of nodes forming clusters $N_{clus} (\bullet)$ as a function of the coupling strength 
$\varepsilon$.  The network structure are same as for the Fig.~(2) and quantities
are plotted for the average over 20 realization of random sets of the 
initial conditions.}
\label{NoClus}
\end{figure}

Fig.~(\ref{InterIntraTime}) shows that after initial turbulent regime where none of the node 
synchronizes , nodes form synchronized clusters. For the weak coupling ($\varepsilon \sim 0.2$) the 
coupled dynamics governed by Eq.~\ref{cml} shows partial synchronized state corresponding to the 
many dynamical clusters with $f_{inter}$ and $f_{intra}$ both being nonzero. Subfigure (a) is 
plotted for the initial network (Fig.~\ref{FigNet}(a)), subfigure (b) is plotted for the rewiring 
stage $n=5$, which means that 5 nodes from each group are rewired. These nodes break the 
connections with their communities and make new connections with the nodes of the other 
communities.

Fig.~(\ref{NoClus}) plots the fraction of number of clusters ($M_{clus}$) and the fraction of number 
of nodes forming clusters $N_{clus}$.  The first measure $M_{clus}$ counts an isolated node  
as a separate cluster, and the second measure $N_{clus}$ counts only those nodes  
which are the part of a cluster and are not isolated. It is clear from the subfigures
that for each rewiring step almost all the nodes form cluster after some coupling strength 
$\varepsilon > 0.2 $ value. Note that for the lower coupling strength region, in general, nodes form 
phase-synchronized cluster, as defined in the previous section, whereas for the stronger coupling
strength region $\varepsilon > 0.5$ they form exact synchronized clusters. For
our investigations it does not matter whether nodes are phase-synchronized or exact synchronized, as
long as large number of nodes form the clusters we get relevant information
regarding dynamical cluster and network structure. It is seen from all the 
subfigures of Fig.~(\ref{InterIntraTime}), that the nature of synchronized clusters is very 
much similar in the weaker coupling strength regime ($\varepsilon \sim 0.2$)
for all rewiring steps, with (a) and (f) being the extreme cases of the network 
structure. Hence, for our study the weaker 
coupling strength region is not very interesting as in this region one does not seem to 
get relation between dynamical cluster and network structure.  In order to get 
insight about the relation between dynamical clusters and the network structure we 
concentrate on the coupling strength region $\varepsilon > 0.5$. Coupled dynamics 
Eq.~(\ref{cml}) gives interesting dynamical clusters in this coupling strength region, where 
connections between the nodes of dynamical clusters differ as structure of 
the network is varied. In the following, the behavior of coupled dynamics and nature of 
synchronized clusters for the various coupling strengths in this region are discussed.

\begin{figure}
\centerline{\includegraphics[width=7.5cm,height=5cm]{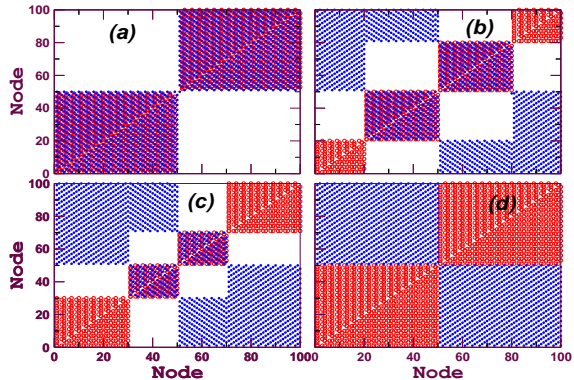}}
\caption{(Color online) Node vs node diagrams illustrating the behavior of nodes
at different step of rewiring. After an initial transient synchronized clusters
are studied. 
The solid circles ($\bullet$) show that the two corresponding nodes
are coupled and the open circles ($\circ$) show that the corresponding nodes are (phase) synchronized. In each case the node numbers are
reorganized so that nodes belonging to the same cluster are numbered consecutively.
All plots are for coupling strength $\varepsilon=1$.
(a) for the network having two complete subgraphs with 
$N_1 = N_2 = 50$ (b) network at the rewiring step $n= 20$ (c) for $n = 30$ 
and (d) for $n = 50$ which leads to the complete bipartite network. }
\label{NodeEps10}
\end{figure}

Fig.~(\ref{NodeEps10}) plots the synchronized clusters for different networks 
generated with the above rewiring strategy. The figures are plotted for the coupling
strength $\varepsilon = 1$. For this coupling strength the nodes form exact synchronized
cluster rather than the phase-synchronized cluster which is the case for weaker
coupling strengths. Fig.~(\ref{NodeEps10})(a) is for the rewiring step 
$n=0$, when nodes in the network form two complete subgraphs. The dynamical behavior of 
these nodes show synchronization with two clusters, and these clusters are of self-organized 
type; i.e. nodes belonging to the structural clusters form dynamical clusters. 
Fig.~(\ref{NodeEps10})(b) is node-node plot for the network at step $n=20$, i.e. 
connections of 20 nodes from each group are rewired such that the 20 nodes from the 
group $G_1$ ($G_2$) get connected with the 30 nodes of the group $G_2$ ($G_1$). At 
this stage mixed dynamical clusters are formed. Two small groups of nodes 
which are rewired to the different groups loose the synchronization with the nodes in 
their respective group and are synchronized independently forming two separate 
clusters. These two clusters, first and fourth cluster from the bottom, are therefore of driven 
type. Rest of the nodes, which remain fully connected inside their respective groups, 
remain self-organized type. 
\begin{figure} 
\centerline{\includegraphics[width=7.5cm,height=5cm]{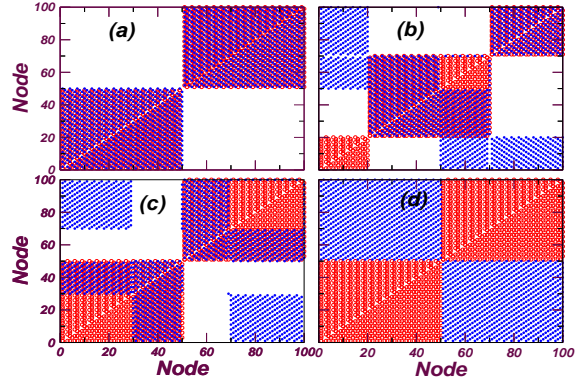} }
\caption{(Color online) Node vs node diagram illustrating the behavior of nodes (same as Fig.~(4)), 
for the coupling strength $\varepsilon = 0.9$. Network structure remains same as of the Fig.~(4).
Node numbers are reorganized so that nodes belonging to the same dynamical 
cluster are numbered consecutively.}
\label{NodeEps09} 
\end{figure}
Fig.~(\ref{NodeEps09}) shows dynamical clusters at the various stage of rewiring
for the coupling strength value $\varepsilon=0.9$. Subfigure (a) shows the two self-organized clusters for the initial network (Fig.~\ref{FigNet}). Subfigures (b), (c) and (d) are plotted
for the $n = 20, 30$ and $50$ rewiring steps respectively. For $n=0$ and $n=50$, self-organized and 
driven clusters respectively are the stable configurations, thus coupled dynamics lead to these
for both the coupling strengths. For intermediate rewiring states dynamical clusters
vary with the value of coupling strengths.
Node numbers are
reorganized so that nodes belonging to the same cluster are numbered consecutively.
\begin{figure}
\centerline{\includegraphics[width=5cm]{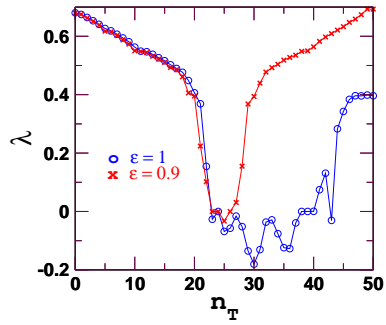}}
\caption{(Color online) Largest Lyapunov exponent $\lambda$ as a function of the rewiring step
$n$. Figures are plotted for two different coupling strengths $\varepsilon = 0.9$
and $\varepsilon = 1$. Lyapunov exponents are calculated for the average over 20 
realization of the random set of initial conditions.}
\label{LyaNet}
\end{figure}
Fig.~(\ref{LyaNet}) plots largest Lyapunov exponent as a function of the rewiring
step $n$, $n=0$ corresponding to the original network (Fig.~\ref{FigNet}(a)).
At each rewiring step one node from each group is rewired. After $n = N_1 = N_2 =50$,
the network becomes like Fig.~(\ref{FigNet})(c). After fixing the coupling strength,
Lyapunov exponent $\lambda$ is: calculated for the coupled dynamics (Eq.~\ref{cml}) at each
step of the rewiring. Results for $\varepsilon = 0.9$ and
$\varepsilon = 1$ show that coupled dynamics may lie on the periodic ($\lambda < 0$),
quasiperiodic ($\lambda \sim 0$) or on the chaotic ($\lambda > 0$) attractor, but
the nature of the coupled dynamics governed by the Eq.~\ref{cml}
remains as shown by the Fig.~(\ref{NodeEps10}) and
\ref{NodeEps09}, depending upon $n$. 

\section{Lyapunov function analysis}
We define Lyapunov function for any pair of nodes $i$ and $j$ as 
\cite{Lyapunov1,Lyapunov2}
\begin{equation}
V_{ij}(t) = (x_i(t) - x_j(t) )^2
\end{equation}
Clearly, $V_{ij}(t) \geq 0$ and the equality holds only when the nodes $i$ and
$j$ are exactly synchronized. For the asymptotic global stability of the
synchronized state in a region, Lyapunov function should satisfy the following
condition in that region:
\begin{equation}
\frac{V(t+1)}{V(t)} < 1 \nonumber
\label{LyaFunc1}
\end{equation}
For the global synchronous
state ($x_i(t) = x_j(t), \forall i,j$), we can write the condition for two synchronized 
clusters in the self-organized state (i.e.
$x_{i_1}(t) = x_{i_2}(t) ;\, \, \forall i_1, i_2 \in G_1$ and 
$x_{j_1}(t) = x_{j_2}(t) ;\, \, \forall j_1,j_2 \in G_2$), as;
\begin{eqnarray}
V_{i_1 i_2}(t+1) &=& \left[ (1-\varepsilon)[f(x_{i_1}(t)) - f(x_{i_2}(t))] \right. \nonumber \\
& & \left. - \frac{\varepsilon}{N_{1}-1} [f(x_{i_1}(t)) - f(x_{i_2}(t))] \right] 
\label{Lya-s-syn}
\end{eqnarray}
Using Eq.~\ref{LyaFunc1} and the above equation for coupled dynamics (Eq.~\ref{cml}),
we get the coupling strength region for which 
the synchronized clusters state is stable;
\begin{equation}
\frac{N1-1}{N1} (1 - \frac{1}{\mu}) < \varepsilon \le 1
\end{equation}

With the rewiring, say at rewiring step $n=\nu$, we get $\nu$ nodes from each
group rewired such that now there are four different types of nodes.
One can quickly see by using
Lyapunov function test that the following synchronized clusters state
would be stable ;\\
1) Cluster $one (S_1)$ with the nodes which remain unwired in the group $G_1$,
synchronized dynamics of nodes in this cluster is $x_i(t) = X_1(t)$.\\
2) cluster $two (S_2)$ with nodes from group $G_1$ which get disconnected 
with the other nodes in $G_1$ and get connected with all the nodes in $G_2$,
$x_i = X_2(t)$.\\
(3) cluster $three (S_3)$ with nodes which remain in group $G_2$, 
$x_i = X_3(t)$.\\
(4) cluster $four (S_4)$, rewired nodes in $G_2$, $x_i(t) = X_4(t)$.\\
Lyapunov function for the nodes in the cluster $one$ (and for the cluster
$three$) remains same as Eq.\ref{Lya-s-syn}, because coupling of the nodes
in this group to the groups $three$ and $four$ ($one$ and $two$) cancel out.
\begin{figure}
\centerline{\includegraphics[width=7cm,height=3.5cm]{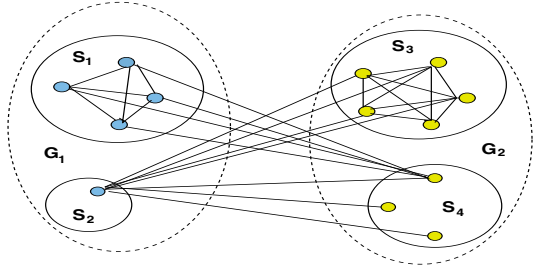}}
\caption{Dynamical clusters at a rewiring state $n_t$. $S_1, S_2, S_3$ and $S_4$
are different dynamical clusters as described in the text. $G_1$ and $G_2$ are the dynamical
clusters for the initial network $n=0$.}
\label{FigLyaFuncNet}
\end{figure}
Lyapunov function for the nodes in group $S_2$ ($S_4$) can be written as;  
\begin{equation}
V_{ij}(t+1) = \left[ (1-\varepsilon) (f(x_i(t)) - f(x_j(t))) \right]^2 \nonumber
\end{equation}
This equation is very simple because the coupling terms for the nodes $i$ 
and $j$ are same and hence they cancel out. Condition for this state to be
stable is given as;
\begin{eqnarray}
\frac{V_{ij}(t+1)}{V_{ij}(t)} &=& (1-\epsilon)^2 \left[ f^{\prime}(x_i(t))
+ \frac{x_i(t) - x_j(t)}{2} f^{\prime \prime}(x_j(t)) \right. \nonumber \\
& & \left. + O((x_i(t)-x_j(t))^2) \right]^2 < 1\nonumber
\end{eqnarray}
For $f(x) = \mu x (1-x)$ and using $0 \le x_i(t) + x_j(t) \le 2$ and condition for
achieving synchronized state, we get the range of coupling strength
as ($(\mu-1)/\mu \le \varepsilon \le 1$) for which the 
synchronized cluster $S_2$ (as shown in the Fig.~(\ref{FigLyaFuncNet}) is stable.
Similarly we can write the condition for the nodes in the group $G_2$ which
are rewired and make synchronized cluster as $S_4$, and nodes which make
synchronized cluster as $S_3$.

\section{Conclusions and Discussions}

In order to explore the relations between dynamical clusters and network clusters we study 
coupled maps on various networks generated with a simple rewiring strategy. Starting with a 
network having two complete subgraphs, nodes are rewired at each step such that 
after some rewiring steps we get a complete bi-partite network. Rewiring strategy is adopted such
that average degree of the networks at each rewiring step remains of the order of $N$. Smaller
coupling strength values show phase synchronized clusters of dominant driven type, whereas
larger coupling strength region show different mechanisms of synchronized clusters formation
depending upon underlying network architecture, and hence provide insight into role of network
architecture in the coherent behavior of the associated dynamical units.
 Coupled dynamics form self-organized clusters for the network having two complete subgraphs, and 
form driven clusters for the bi-partite case. At intermediate steps, the nodes receiving similar 
input form dynamical clusters, and these clusters could be self-organized type, driven type or 
mixed type depending upon which connection environment they belong to. Lyapunov function analysis 
shows that for the driven synchronization if any two nodes have similar coupling architecture, the 
difference of the dynamical variables for these two nodes cancel out. Whereas for the 
self-organized synchronization, the coupling term corresponding to the direct coupling between the 
nodes do not cancel out, and other couplings which are common to both the nodes cancel out. The 

Ref.~\cite{SJ-analytic} shows that there are two mechanisms of cluster formation in networks, and 
using the global stability analysis it provides arguments behind the mechanisms by taking globally 
coupled and complete bipartite as the extreme cases. Here we show that there is a gradual transition 
from the self-organized to the driven behavior, as the underlying network is rewired from the two clusters to the 
bipartite network. The small coupling values do not show any impact of the structural changes on the 
mechanism of the synchronization, whereas large coupling values show significant signature of the 
underlying structure on the synchronized clusters. The dense networks ($N_c \sim N^2$) considered 
here yield all the nodes forming clusters at each rewiring stage, and 
clusters are of different types depending upon the underlying network structure at that particular 
rewiring step. Through extensive numerical simulations, we show that the nodes, getting similar coupling
environment, are synchronized irrespective of whether they are connected or not. At the intermediate 
rewiring steps and for the higher coupling values, few clusters are mixed type, and few clusters are of the 
ideal driven or of the ideal self-organized type. All the nodes in each cluster receive similar 
coupling environment.  Using the Lyapunov function analysis, we get a clear picture of the 
synchronization for each cluster individually. The stability condition for the synchronization of 
the nodes in clusters $S_2$ and $S_4$ leads to the reasoning valid for the ideal driven
cluster. Also, stability range 
written for the dynamical clusters $S_1$ and $S_3$ matches with that of the ideal self-organized 
cluster because of the very simplistic underlying network model we are using, where any terms 
outside these cluster cancel out leading to the condition shown by the globally coupled network in 
the synchronized state.

Though the network architecture and rewiring strategies considered here are very simple, whereas 
real world networks have complicated random structure or have complicated rewiring or connection 
evolution strategies \cite{REA-Chaos}, the results presented here shows direct impact of network 
connection architecture on the evolution of dynamical units. The main aim of the analysis here is 
to emphasize that the dynamical clusters {\it do have} information about the structure, but the 
formation of driven clusters reveal that the nodes in the dynamical clusters are not always the 
nodes which are connected, rather the nodes which have similar environment show coherent behavior 
and form a dynamical cluster. For a given network, depending upon the coupling strengths there 
could be various possible states which form stable configurations of the dynamical clusters.  
Results presented in the paper indicate that synchronous behavior of the dynamical units 
interacting with each other might be receiving the similar inputs, for instance synchronous firing 
of neurons \cite{syn-neuro} in a cortex may be because of the couplings among the neurons, or may 
be because of getting similar synaptic inputs from the neurons of different cortical areas. In the 
continuation of previous work \cite{SJ-prl1,SJ-pre1,Changsong2011}, the results presented here 
strengthens understanding about the possible relations between the dynamical behavior and the 
underlying connections architecture. Future investigations would involve more general network 
rewiring or evolution scheme, as well as effect of delays \cite{delay} on the phenomenon of driven 
synchronization.

\section{Acknowledgement} SJ thanks DST for financial support.

\end{document}